\documentclass[pre,floatfix,superscriptaddress,twocolumn,showpacs]{revtex4-1}
\usepackage[english]{babel}
\usepackage{graphicx}
\usepackage[table]{xcolor}
\usepackage{amsmath,amssymb,amsfonts}
\usepackage{mathtools}
\usepackage{bm}
\usepackage{color}
\usepackage{multirow}
\usepackage{url}
\usepackage{outlines}
\normalsize
\usepackage{booktabs}
\usepackage{xparse}
\usepackage{float}
\usepackage{layouts}

\usepackage{tikz}
\usepackage{graphicx}
\usepackage{import}
\usetikzlibrary{positioning}
\usetikzlibrary{arrows.meta}

\usepackage[acronym]{glossaries}

\newcommand{\dmin}{D^2_\mathrm{min}}

\usepackage{hyperref}
\usepackage{tabularx}
\renewcommand{\refeq}[1]{eq.~(\ref{#1})}

\NewDocumentCommand{\reffig}{m o}{%
  Fig.~\ref{#1}%
  \IfValueT{#2}{#2}%
}

\newacronym{msd}{MSD}{mean square displacement}
\newacronym{spv}{SPV}{self-propelled Voronoi}
\newacronym{cpm}{CPM}{cellular Potts model}
\newacronym{emt}{EMT}{epithelial-mesenchymal transition}
\newacronym{ecm}{ECM}{extracellular matrix}
\newacronym[shortplural={MCS}]{mcs}{MCS}{Monte-Carlo sweep}
\newacronym[shortplural={COM},longplural={centres of mass}]{com}{COM}{centre of mass}
\newacronym{fft}{FFT}{fast Fourier transform}
\newacronym{mape}{MAPE}{mean absolute percentage error}

\begin{document}

\title{Nuclear mechanics controls the temporal dynamics of cell unjamming}
\author{Leon Hillmann}
\affiliation{Department of Applied Physics, Eindhoven University of Technology, P.O.~Box 513, 5600 MB Eindhoven, The Netherlands}
\affiliation{Department of Mathematics and Computer Sciences, Eindhoven University of Technology, P.O.~Box 513, 5600 MB Eindhoven, The Netherlands}
\affiliation{Department of Biomedical Engineering, Eindhoven University of Technology, P.O.~Box 513, 5600 MB Eindhoven, The Netherlands}

\author{Quirine J.~S.~Braat}
\thanks{Contributed equally}
\affiliation{Department of Applied Physics, Eindhoven University of Technology, P.O.~Box 513, 5600 MB Eindhoven, The Netherlands}

\author{Pablo Gottheil}
\thanks{Contributed equally}
\affiliation{Peter Debye Institute for Soft Matter Physics, University of Leipzig,
Linnéstraße 5, Leipzig, 04103, Saxony, Germany}

\author{Eliane Blauth}
\affiliation{Peter Debye Institute for Soft Matter Physics, University of Leipzig,
Linnéstraße 5, Leipzig, 04103, Saxony, Germany}

\author{Anne Marie Scholz}
\affiliation{Peter Debye Institute for Soft Matter Physics, University of Leipzig,
Linnéstraße 5, Leipzig, 04103, Saxony, Germany}

\author{Kolya M. Lettl}
\affiliation{Peter Debye Institute for Soft Matter Physics, University of Leipzig,
Linnéstraße 5, Leipzig, 04103, Saxony, Germany}

\author{Jürgen Lippoldt}
\affiliation{Peter Debye Institute for Soft Matter Physics, University of Leipzig,
Linnéstraße 5, Leipzig, 04103, Saxony, Germany}

\author{Pieta C.~M.~Wielstra}
\affiliation{Department of Applied Physics, Eindhoven University of Technology, P.O.~Box 513, 5600 MB Eindhoven, The Netherlands}

\author{Sibylle Hess}
\affiliation{Department of Mathematics and Computer Sciences, Eindhoven University of Technology, P.O.~Box 513, 5600 MB Eindhoven, The Netherlands}

\author{Mitko Veta}
\affiliation{Department of Biomedical Engineering, Eindhoven University of Technology, P.O.~Box 513, 5600 MB Eindhoven, The Netherlands}

\author{Josef A. Käs}
\email{jkaes@uni-leipzig.de}
\affiliation{Peter Debye Institute for Soft Matter Physics, University of Leipzig,
Linnéstraße 5, Leipzig, 04103, Saxony, Germany}

\author{Liesbeth M.~C.~Janssen}
\email{l.m.c.janssen@tue.nl}
\affiliation{Department of Applied Physics, Eindhoven University of Technology, P.O.~Box 513, 5600 MB Eindhoven, The Netherlands}
\affiliation{Institute for Complex Molecular Systems, Eindhoven University of Technology, P.O.~Box 513, 5600 MB Eindhoven, The Netherlands}

\date\today

\begin{abstract}
    Cell unjamming in dense tissues is a complex but essential process in
    embryogenesis and cancer metastasis. Increasing evidence suggests that
    nuclear mechanics and density effects play a vital role in collective cell
    unjamming. However, state-of-the-art cell-shape-based theories fail to
    include nuclear and density effects, while computer models featuring
    rigid nuclei disagree with experimental observations of elongated nuclei
    promoting unjamming. Here, we introduce a computational model of confluent
    cells with explicitly deformable nuclei to study the dynamics of cell unjamming. Our
    simulations show an unjamming transition controlled by nuclear size and
    shape, reconciling conflicting theories of density-driven versus
    shape-driven mechanisms. We predict general relations connecting cellular and
    nuclear shape to collective cell motion, verified experimentally in distinct monolayers of MCF-10A
    and MDA-MB-436 breast cells, with striking accuracy. Our work establishes a
    rational connection between nuclear mechanics and tissue-scale rigidity
    transitions, highlighting the nucleus's key role in collective cell
    unjamming.
\end{abstract}
\maketitle

\section{Introduction}

The ability of cells to collectively rearrange underlies both tissue function
and dysfunction across embryonic development, wound healing, and cancer
progression. These rearrangements have recently been
linked to a so-called jamming-unjamming transition where tissues
switch between disordered solid-like and fluid-like
states~\cite{friedl_collective_2009, park_unjamming_2015}. Increasing evidence
suggests that the physics of jamming and glass formation, long established for
inanimate liquids and granular materials, also applies to living
tissues~\cite{atia_are_2021, mitchel_primary_2020,
bi_density-independent_2015, angelini_glass-like_2011}. In cancer, phenotypic
fluidity transitions such as the \gls{emt} have long been associated with
increased invasiveness and metastatic
potential~\cite{francou_epithelial--mesenchymal_2020,
nishiokaSNAILInducesEpithelialtomesenchymal2010}. More recently, tissue-level
unjamming has been linked to the onset of collective cell motility,
a prerequisite for metastasis~\cite{oswald_jamming_2017, gottheil_state_2023,
grosser_cell_2021}. The quintessential importance of unjamming for tumour
progression has recently been demonstrated by the observation that unjamming in
the primary tumour serves as a significant prognostic marker for metastatic
risk~\cite{gottheil_state_2023}. 

While unjamming has been shown to regulate tissue fluidity, the mechanisms that control this transition on the cellular
and tissue level remain unclear. Pioneering studies on confluent monolayers have shown that tissue fluidisation depends on the average cell shape \cite{bi_density-independent_2015,bi_motility-driven_2016}, but more recent experiments indicate that unjamming also correlates with nuclear
shape~\cite{gottheil_state_2023, yuNuclearPackingSets2026}. The role of the nucleus -- the largest organelle in the cell -- remains elusive however, since the only studies to date on nuclear effects in cell jamming reveal a clear discrepancy: 
Models for morphogenesis in
zebrafish predict that cells with elongated, infinitely stiff nuclei are
more prone to jamming~\cite{kim_nuclear_2024}, whereas in breast cancer
explants, elongated cells and nuclei promote
unjamming~\cite{gottheil_state_2023}. Furthermore, given that cancer cell nuclei are deformable rather than rigid~\cite{fuhs_rigid_2022}, this suggests that the softness of nuclei may also play a vital, but previously overlooked, role in understanding and predicting cell jamming.    

Previous studies of cell jamming using vertex, \gls{spv}, and cellular Potts models
 treat cells as homogeneous mechanical objects, omitting subcellular
structures such as the nucleus, thus leaving their role in tissue mechanics poorly
understood~\cite{bi_motility-driven_2016, bi_density-independent_2015,
devanny_signatures_2023}. Since cells have to squeeze by each other to unjam,
cell deformability is a key determinant in cell unjamming. Indeed, it has been shown
that cancer cell softening promotes unjamming and
metastasis~\cite{fuhs_rigid_2022, guck_optical_2005}. Those studies, however,
focused on the stiffness of the cytoskeleton. Despite direct experimental
evidence that cell nuclei contribute to tissue mechanics~\cite{xie_effect_2024, biswasConservedNucleocytoplasmicDensity2025}
and strongly influence cancer cell unjamming~\cite{gottheil_state_2023}, current
models have not considered deformable nuclei as an essential mechanical element.

\begin{figure*}[tbh]
    \includegraphics[width=\textwidth]{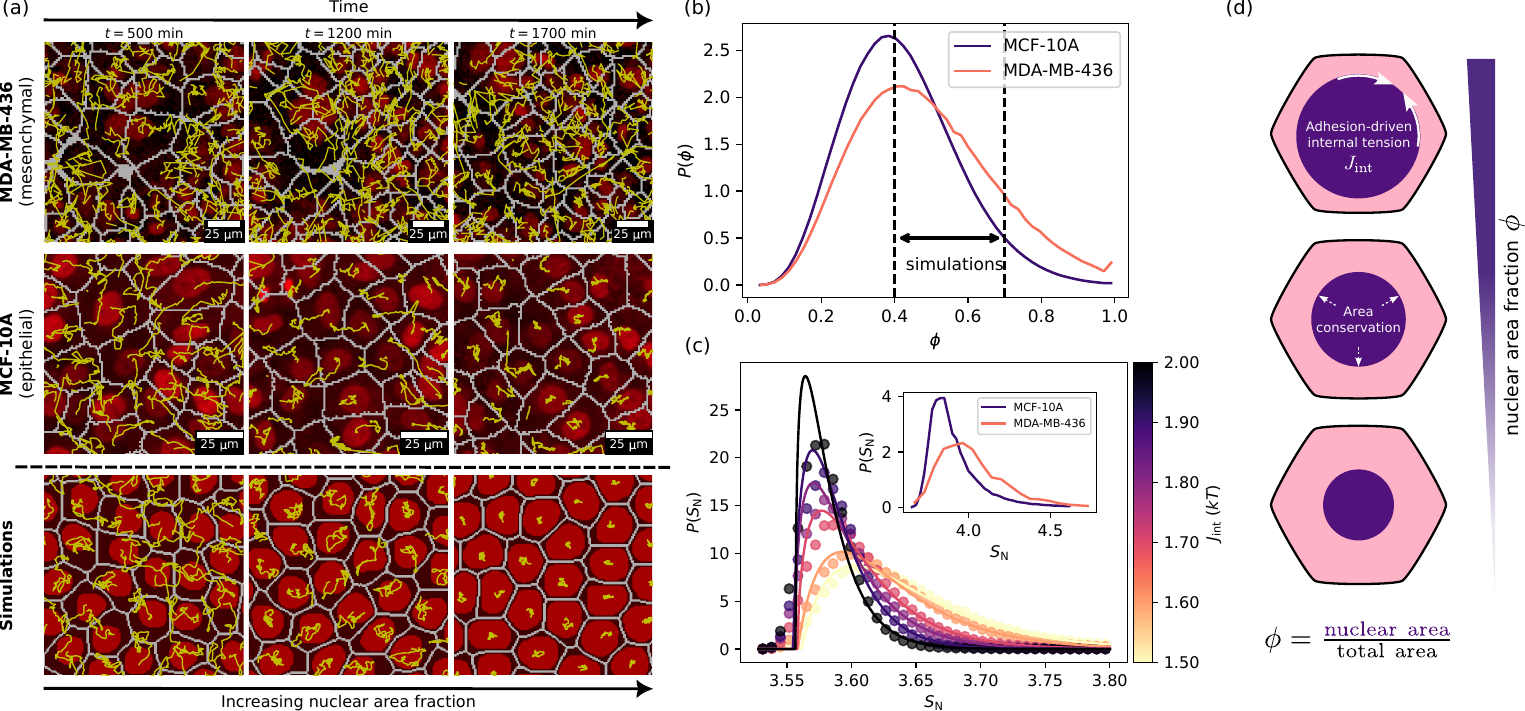}
    \caption{\textbf{Nuclear area fraction and stiffness control cell morphologies
    and cell dynamics.} \textbf{a},~2D layers of MCF-10A and MDA-MB-436 at 500, 1200, and 1700 minutes after the start of experiments compared to simulations at different nuclear area fractions. Top: 2D layers of mesenchymal MDA-MB-436
    cells with trajectories over the course of 360 minutes. The trajectories show that tissues remain unjammed.
    Middle: Layers of epithelial MCF-10A cells with trajectories over the course of 720 minutes. As time progresses,
    layers proliferate and cell density (nuclear area fraction) increases,
    causing a jamming transition. Bottom: Simulation snapshots of confluent
    tissues featuring deformable nuclei with trajectories shown in yellow.
    Increasing nuclear packing causes a jamming transition. The scale bars
    amount to $25\,\mu \mathrm{m}$. \textbf{b},~Histogram of experimentally
    observed nuclear area fractions. Nuclear area fractions in MDA-MB-436 are
    slightly higher than those found in MCF-10A cells. We choose to study the
    far tail of the distribution in simulations. \textbf{c},~Histograms of
    simulated nuclear shapes for different nuclear stiffnesses. Solid lines are
    fitted gamma distributions, which have previously been observed in similar
    contexts \cite{atia_geometric_2018}. Inset: Histogram of nuclear shapes
    observed in experiments. Shape distributions in MDA-MB-436 cells are broader
    and exhibit a longer tail, owed to their softer nuclei. \textbf{d},~Our
    computational model considers two control parameters: Nuclear area fraction $\phi$,
    determined by the nucleus's target area in the \gls{cpm}'s area conservation
    term (see section \ref{methods:model}), and nuclear stiffness, controlled by
    the adhesion parameter $J_\mathrm{int}$.
    }
    \label{fig:1-sims-expts}
\end{figure*}
Here, we introduce a
new \gls{cpm} that captures the viscoelastic interplay
between the cytoplasm and deformable nuclei. Our
computational model accurately captures unjamming in monolayers of MCF-10A epithelial and
MDA-MB-436 mesenchymal cells. Although these cell lines represent two
phenotypes on opposite ends of the \gls{emt}, we show that the dynamics of both
obey the same physical mechanisms. We introduce a universal
relation connecting nuclear shape and packing density with cell
diffusion. Existing theories have not yet resolved whether jamming is driven by
cell density or shape~\cite{oswald_jamming_2017}. Our model reconciles these
two views: We find that, while cell
shape determines jamming phenomenologically, nuclear density constrains
possible cell shapes and thereby regulates the transition at the subcellular scale. Given the immanent
importance of the nucleus for tumour biology –- nuclear
deformation contributes to mechanosignaling and nucleus grading is an important
diagnosic tool~\cite{lammerding_mechanics_2011,
bloomHistologicalGradingPrognosis1957} –- our model can also advance our
understanding of pathological processes and contribute to new
physics-informed prognostic markers for cancer metastasis.

\section{Results}

\subsection{Larger, stiffer nuclei promote jamming}

To address the role of the nucleus in unjamming, we investigate how nuclear
properties affect collective dynamics of two cell lines: the non-transformed
mammary epithelial MCF-10A and the mesenchymal MDA-MB-436 from invasive breast
cancer. MCF-10A cells proliferate, which increases cell density and results in a
jamming transition (\reffig{fig:1-sims-expts}[a]). In contrast, the number
density of unjammed MDA-MB-436 cells remains approximately constant. Compared with
MCF-10A, MDA-MB-436 cells lack cortical tension and contain larger, more
deformable nuclei~\cite{fuhs_rigid_2022} (\reffig{fig:1-sims-expts}[b-c]).

To examine nuclear mechanics during jamming, we develop a
simulation model in which we control nuclear size and stiffness. Our model allows
us to probe different state points throughout the transition and thus offers clear insights into the nuclear
mechanisms affecting jamming. We focus on two-dimensional
systems to compare with existing, nucleus-free models of cell jamming
~\cite{bi_motility-driven_2016, devanny_signatures_2023,
chiang2016glass, sadhukhan2021theory}, but our model may be straightforwardly
generalized to three dimensions. Inspired by the single-cell model by \citet{scianna_cellular_2021}, we introduce
a cellular Potts model of a confluent cell layer with
elastic nuclei. Cell migration is modelled by an active force with constant
magnitude and randomly diffusing polarity~\cite{lecuit_force_2011,
anon_cell_2012, henkes_dense_2020}. Motivated by the idea that the nucleus can
act as a mechanical obstacle to migration~\cite{mcgregor_squish_2016,
wolf_physical_2013, fuhs_rigid_2022}, we focus on two parameters: (i) nuclear
area fraction $\phi$, and (ii) nuclear stiffness $J_\mathrm{int}$
(\reffig{fig:1-sims-expts}[d]).

We define nuclear area fraction as $\phi = A_\mathrm{N} / A_\mathrm{C}$ where
$A_\mathrm{C}$ and $A_\mathrm{N}$ are the target areas of the cell and the
nucleus, respectively. The cell area fluctuates around a
fixed $A_\mathrm{C}$, while the nuclear target area $A_\mathrm{N}$ is varied to
control the average nuclear area fraction (see section \ref{methods:model}). Our
experimental monolayers exhibit distributions of $\phi$
with a mean of $\bar{\phi} = 0.42 \pm 0.15$ for MCF-10A cells and
a slightly larger packing fraction of $\bar{\phi} = 0.48 \pm 0.18$ for
MDA-MB-436 (\reffig{fig:1-sims-expts}[b]). In
simulations, we cover the right flank of that distribution by varying $\phi$
between $0.4$ and $0.7$, since we expect the nuclear effects to become most
dominant at large $\phi$.

Nuclear stiffness is controlled by the internal contact energy $J_\mathrm{int}$
between the nucleus and its surrounding cytoplasm (see section
\ref{methods:model}). $J_\mathrm{int}$ can be directly related to
the interfacial tension of the nuclear envelope \cite{magno_biophysical_2015}.
Higher values of $J_\mathrm{int}$ correspond to increased stiffness and hence
yield more compact, round nuclei. Following previous cell-shape-based studies
\cite{bi_density-independent_2015, park_unjamming_2015, grosser_cell_2021,
gottheil_state_2023}, we quantify these differences in nuclear shape with the dimensionless nuclear shape index
$S_\mathrm{N} = \left\langle\frac{P_\mathrm{N}}{\sqrt{A_\mathrm{N}}}
\right\rangle$. Here, $P_\mathrm{N}$ and $A_\mathrm{N}$ are the nuclear
perimeter and area, respectively, and the brackets denote an ensemble average.
Simulations with $J_\mathrm{int}$
between $1.5\, k_\mathrm{B}T$ and $2.0\, k_\mathrm{B}T$ (in units of thermal
energy $k_\mathrm{B}T = 1$) qualitatively mirror the distributions of
nuclear shape indices in our experiments (\reffig{fig:1-sims-expts}[c]).
The softer
MDA-MB-436 nuclei exhibit a broader shape distribution compared to the more
rigid MCF-10A nuclei. In simulations, we choose narrower
distributions to study the effects of specific nuclear
conditions on tissue fluidity. Our simulated nuclei are also more circular than
those observed in experiments (see inset in \reffig{fig:1-sims-expts}[c]), which
indicates higher stiffness and thus amplifies the effect of the
nucleus on cell jamming in our simulations.

We systematically vary $\phi$ and $J_\mathrm{int}$ to quantify their
influence on collective dynamics. We measure common
indicators of jamming in glassy systems~\cite{bi_density-independent_2015,
bi_motility-driven_2016}: the rate of T1 transitions, the average \gls{msd}, and
the corresponding long-time diffusion constant $D$. In experiments, we quantify
dynamics by the closely related cage-relative non-affine displacement $\dmin$
(see section \ref{methods:experimental-setup}). 
Higher $D$ or $\dmin$ indicate higher mobility (unjammed states), while lower
values correspond to solid-like, jammed layers with a lower rate of T1
transitions.

\begin{figure*}[tbh]
    \includegraphics[width=\textwidth]{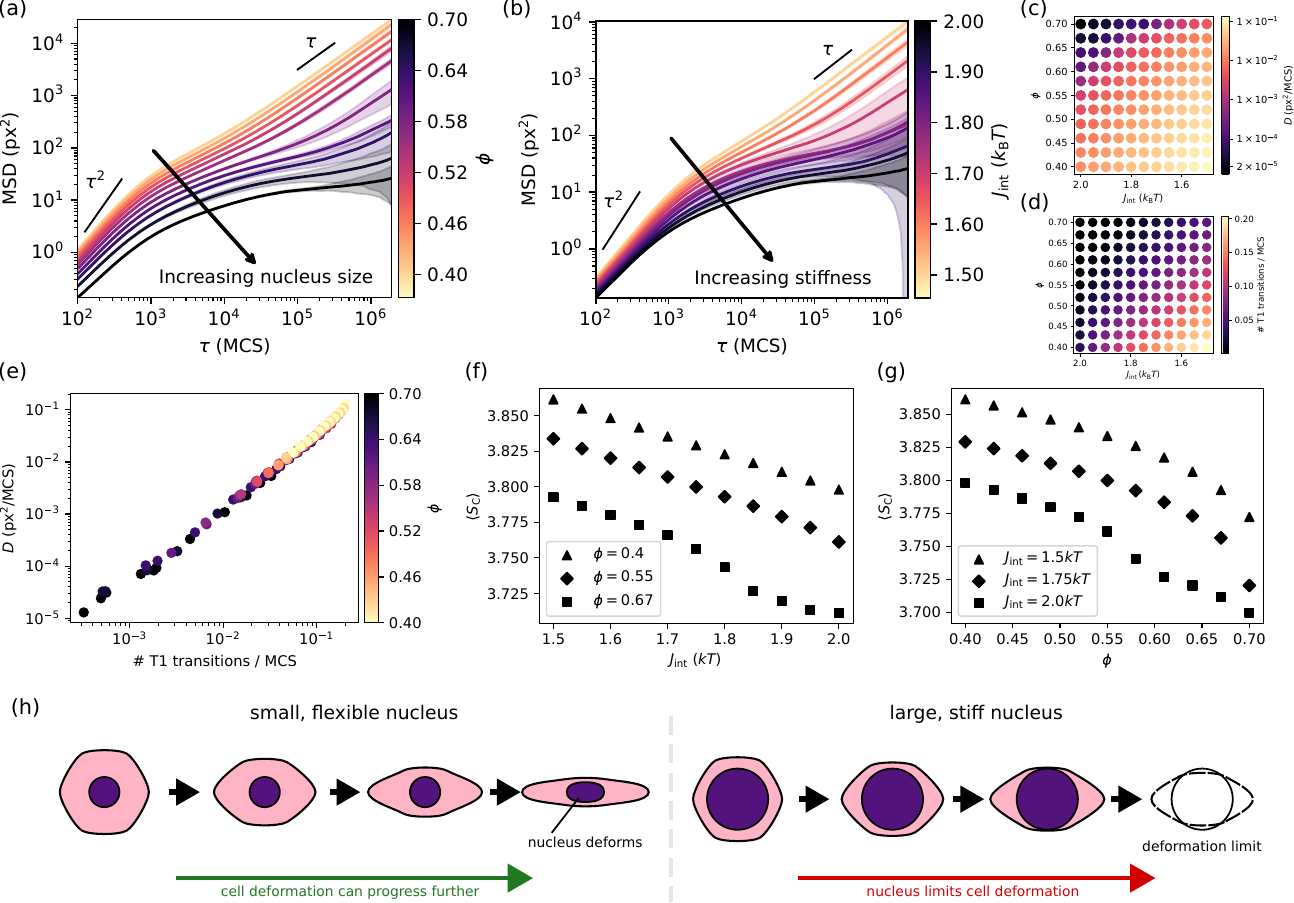}
    \caption{
    \textbf{Larger and stiffer nuclei limit diffusion.}
    \textbf{a},~\gls{msd} at constant stiffness $J_\mathrm{int} = 2\,k_\mathrm{B}T$ and
    increasing area fractions.
    \textbf{b},~\gls{msd} at constant area fraction $\phi = 0.67$ and increasing
    stiffness.
    Shaded areas in \textbf{a} and \textbf{b} indicate the standard deviation
    over 10 independent simulation runs. 
    \textbf{c},~Diffusion constants for various nuclear area fractions $\phi$
    and internal contact energies $J_\mathrm{int}$. Higher contact energies
    (nucleus stiffness) and area fractions (nucleus size) lead to a decline in
    the diffusion coefficient. 
    \textbf{d},~Rate of T1 transitions for various nuclera area fractions and
    stiffnesses. The x-axes in \textbf{b} and \textbf{c} are inverted for better
    comparability with \reffig{fig:shape-diffusion-law} (the nuclear shape index
    increases to the right).
    \textbf{e},~Relation between the diffusion constant $D$ and the rate of T1
    transitions in simulations.
    \textbf{f},~Average cell shape in simulations at various
    nuclear stiffnesses $J_\mathrm{int}$. Stiffer nuclei lead to more compact
    cell shapes.
    \textbf{g},~Average cell shape in simulations at various
    nuclear area fractions $\phi$. More densely-packed nuclei lead to more
    compact cell shapes. \textbf{h}, Nuclear properties control the range of
    available cell shapes, schematically shown for a cell subjected to a
    vertical force. Small, flexible nuclei (left) allow for larger deformations
    than larger, stiffer nuclei (right).
    }
    \label{fig:msd_phase_diagram}
\end{figure*}

Our simulation results (\reffig{fig:msd_phase_diagram}[a-c]) show that more densely
packed and stiffer nuclei promote jamming.
We find that nuclear properties control the amount of
caging in the characteristic ballistic-diffusive-subdiffusive
sequence~\cite{bechinger_active_2016, bi_motility-driven_2016} of the cellular
\glspl{msd}.
Increasing either nuclear area fraction or stiffness prolongs
the subdiffusive regime to last for more than $10^5$ \glspl{mcs}, marking the
transition from an unjammed, fluid-like to a jammed, solid-like tissue. This
glass-like transition is further characterised by a sharp decrease in the long
time diffusion constant $D$. Increasing the nuclear area fraction from $\phi = 0.4$
to $\phi = 0.7$ leads to a decrease of $D$ by one
order of magnitude. Likewise, a similar reduction in diffusivity is
observed as $J_\mathrm{int}$ goes from
$1.5\,k_\mathrm{B}T$ to $2\,k_\mathrm{B}T$. Interestingly,
increasing both $J_\mathrm{int}$ and $\phi$ simultaneously yields a
compounded effect: Tissues with the largest, stiffest nuclei exhibit diffusion constants almost four orders of magnitude
lower than those with the smallest and softest nuclei
(\reffig{fig:msd_phase_diagram}[c]). Nuclear area fraction and nuclear
stiffness therefore influence collective dynamics even more strongly than either
of them individually. A similar trend has been found experimentally
by \citet{gottheil_state_2023}, who observed that nuclear volume and shape
correlate with mobility in three dimensional systems. 
In the
following sections, we investigate the mechanism through which
nuclear properties control tissue fluidity.

\subsection{Nuclear properties constrain cell shape}

To study how nuclear area fraction and stiffness control cell dynamics, we
measure the rate of neighbour exchange events through T1
transitions in our simulations (see section~\ref{methods:t1}).  
Figure \ref{fig:msd_phase_diagram}d shows that larger, stiffer
nuclei drastically reduce the rate of T1 rearrangements by almost
three orders of magnitude. The frequency of T1 transitions thus behaves analogously to the long-term diffusion constant discussed above
(\reffig{fig:msd_phase_diagram}[c]). As expected, we observe a direct relation
between $D$ and the rate of T1 transitions (\reffig{fig:msd_phase_diagram}[e]). Mobility in the cell
layer is therefore directly determined by the number of T1 transitions over time.
Nuclear area fraction and stiffness hence control tissue-level jamming and
unjamming by affecting the rate of neighbour exchanges through T1 rearrangements.

\citet{bi_density-independent_2015} recently
linked the rate of T1 transitions to the cell shape index, which is defined (analogously to $S_\mathrm{N}$) as 
$S_\mathrm{C} =
\left\langle\frac{P_\mathrm{C}}{\sqrt{A_\mathrm{C}}}\right\rangle$, where
$P_\mathrm{C}$ and $A_\mathrm{C}$ are a cell's perimeter and area.
Using geometric arguments, Bi \textit{et al.} showed that the energy barrier of T1 transitions in confluent
2D layers vanishes if $S_\mathrm{C} \gtrsim 3.81$~\cite{bi_density-independent_2015}. This
threshold has since been linked to an unjamming transition in various theoretical~\cite{bi_motility-driven_2016, devanny_signatures_2023} and experimental~\cite{park_unjamming_2015, fuhs_rigid_2022} works. The \gls{spv} model
used in those studies however cannot explain how cell nuclei or number
density affect cell shape and dynamics. A new model of cells with infinitely
stiff nuclei~\cite{kim_nuclear_2024} has partially rectified these issues, but
predicts that more elongated nuclei promote unjamming, contrary to the dynamics 
observed in cancer cells~\cite{gottheil_state_2023}. Our \gls{cpm}
provides direct insights into the effects of both -- nuclear density and
deformability -- on emergent cell morphologies. 
We therefore continue by investigating
how nuclear stiffness and area fraction influence cell shape.

We find that stiffer nuclei and larger nuclear area fraction both
lead to rounder cells. The
average cell shape index decreases from $S_\mathrm{C} \approx
3.83$ to $S_\mathrm{C} \approx 3.77$ when increasing nuclear
stiffness $J_\mathrm{int}$ from $1.5\,k_\mathrm{B}T$ to $2.0\,k_\mathrm{B}T$ at
$\phi = 0.55$ (\reffig{fig:msd_phase_diagram}[f]). Similarly, the cell
shape index shifts from $S_\mathrm{C} \approx 3.83$ to $
S_\mathrm{C} \approx 3.73$ upon increasing the nuclear area fraction
$\phi$ from $0.4$ to $0.7$ at $J_\mathrm{int} = 1.75\, k_\mathrm{B}T$
(\reffig{fig:msd_phase_diagram}[g]). Mechanically, there
are two ways for cell deformation. First, cells can deform while
leaving the shape of their nuclei unaltered, a process restricted by the nuclear
area fraction: Larger nuclei limit the room for cells to deform freely, akin to
crowding in density-induced jamming.
Larger nuclear area fractions therefore cause cell shapes to approach the
nuclear shape. Once this nucleus-free mechanism is exhausted, the nucleus itself can deform.
This process directly depends on the stiffness of the nucleus, which is
generally much higher than that of the cell~\cite{mcgregor_squish_2016}. Indeed, stiffer
nuclei require larger external forces to deform, and this property translates to the
shape of the entire cell. Together, these two mechanisms elucidate the pathway of nuclear
jamming: Nuclear size and stiffness control the range of available cell shapes
(\reffig{fig:msd_phase_diagram}[h]). More compact
cell shapes energetically discourage T1 transitions and thus lead to less
mobile, jammed tissues.

\begin{figure*}[thbp]
    \includegraphics[width=\textwidth]{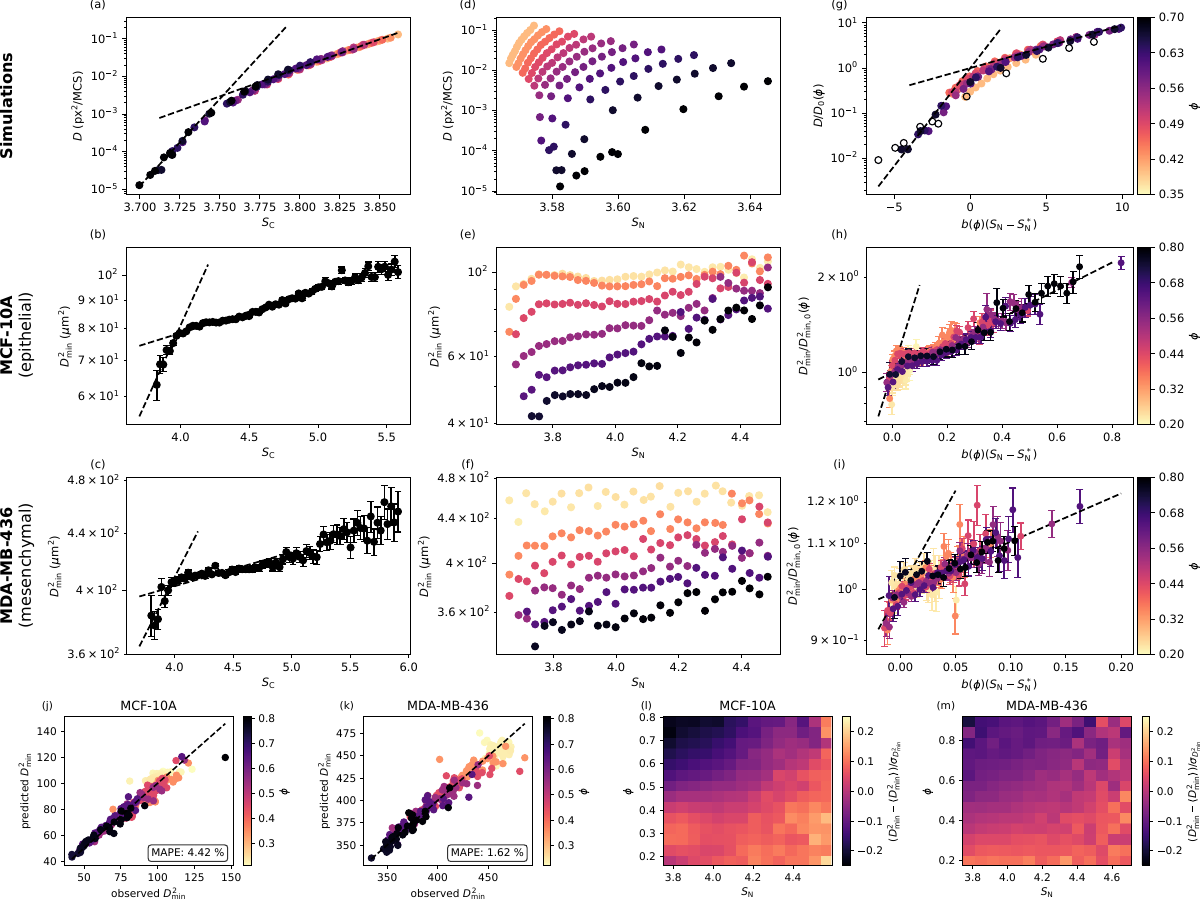}
    \caption{\textbf{Relations linking cellular and nuclear morphologies with mobility.}
    \textbf{a}, \textbf{b}, \textbf{c}, Diffusion constants ($\dmin$ for experiments) exhibit two distinct
    linear regimes when plotted logarithmically against cell shape in
    simulations and experiments with two different cell lines. \textbf{d},
    \textbf{e}, \textbf{f}, Plots of the diffusion coefficient ($\dmin$ for experiments) against nuclear
    shape separated by nuclear area fraction $\phi$. \textbf{g}, \textbf{h},
    \textbf{i}, Rescaling the nuclear shape-diffusion relation collapses data
    from simulations and experiments onto a single master curve. Our simulations
    and both the epithelial and mesenchymal cell lines show the same
    qualitative relation between area fraction, nuclear shape and diffusivity.
    Open symbols in \textbf{g} were excluded from the fit procedure to
    accomplish collapse (see \ref{methods:rescaling}).
    Colours indicate the nuclear area fraction $\phi$ consistently across all
    panels in a row, except in \textbf{b} and \textbf{c}. Error bars show
    the (rescaled) standard error of the mean for experiments in \textbf{b},
    \textbf{c}, \textbf{h}, and \textbf{i}. \textbf{j}, \textbf{k}, $\dmin$
    predicted from our fitted model against $\dmin$ observed in experiments with
    MCF-10A and MDA-MB-436. The dashed line represents $y = x$. The \gls{mape}
    lies below $5\,\%$ for both cell lines. \textbf{l}, \textbf{m}, Relative
    differences in binned $\dmin$ measured in experiments with MCF-10A and
    MDA-MB-436 cells. Colours indicate the difference of each bin average from
    the ensemble-average $\langle \dmin  \rangle$ normalised by the standard
    deviation $\sigma_{\dmin}$. Cells with larger nuclear area fractions and
    rounder nuclei exhibit smaller $\dmin$ than those with smaller nuclear area
    fractions and more elongated nuclei.}
    \label{fig:shape-diffusion-law}
\end{figure*}

A surprising consequence of this mechanism is that cell shape emerges as
an intermediary indicator for nuclear jamming. Our simulation data in
\reffig{fig:shape-diffusion-law}[a] exhibit a quantitative relation between the
cellular shape index and the diffusivity that reaches beyond the formerly
theorised threshold of $S_\mathrm{C} \approx 3.81$. Plotting the long-time
diffusion constant $D$ against the average cell shape index $S_\mathrm{C}$
reveals a single master curve. The graph shows that an
increase from $S_\mathrm{C} = 3.7$ to
$S_\mathrm{C} = 3.8$, i.e.\ an increase in cell elongation, goes along with
a drastic increase of $D$ of roughly three orders of
magnitude. On a logarithmic scale, $D(S_\mathrm{C})$ exhibits a
crossover between two distinct regimes, transitioning from a steeper
to a shallower slope at a shape index of $S^*_\mathrm{C} \approx 3.75$. For
compact cell shapes below $S^*_\mathrm{C}$, small
deformations thus have a relatively large effect on the diffusivity. Beyond this
threshold, more elongated cells still promote mobility, but less strongly.

The same relation between cell shape and mobility can be observed in our
experiments. When plotting $\dmin$ against
cell shape (\reffig{fig:shape-diffusion-law}[b-c]), we recover a relation between
$\dmin$ and $S_\mathrm{C}$ of the same shape as for our simulations. The graphs
again exhibit a transition between two regimes, for MCF-10A cells at $S^*_\mathrm{C} \approx 3.97$ and for MDA-MB-436 cells at $S^*_\mathrm{C} \approx
3.95$.

Strikingly, our simulation data collapse across all examined state
points (\reffig{fig:shape-diffusion-law}[a]). Configurations with the same
emergent cell shape therefore result in the same long-time diffusion constant,
even if the underlying nuclear properties are different. This observation can be
directly explained by the mechanism of nuclear jamming proposed above. The
average cell shape determines the rate of T1 transitions and hence the long-time
diffusion constant, rendering it a universal descriptor of tissue
fluidity. Cell shape alone however does not explain the root cause of jamming in our simulations. The
nucleus constrains possible cell shapes. Its mechanical properties therefore
control which part of the cell shape-diffusion relation is accessible for a
particular cell. Consequently, nuclear morphologies must be included in a full
description of cell jamming and unjamming.

\subsection{A quantitative relation between diffusion and nuclear morphologies}

We now seek a structural order parameter for collective cell dynamics that captures the nuclear control
parameters of the shape diffusion relation. Inspired by
\citet{gottheil_state_2023}, who considered combined cell shape and nuclear
aspect ratio (``CeNuS''), we introduce the average nuclear shape index $S_\mathrm{N}$ as
an order parameter for nuclear morphology. $S_\mathrm{N}$ is directly
related to nuclear stiffness as discussed above. Likewise, nuclear density can
be accounted for with the nuclear area fraction $\phi$. Both quantities are
straightforward to measure from static snapshots.

Plotting $D$ from simulations against the nuclear shape index $S_\mathrm{N}$
(\reffig{fig:shape-diffusion-law}[d]), we observe a clear dependency between
nuclear shape and cell diffusivity, stratified by nuclear area fraction. Small
nuclei do not restrict cells as much, and, therefore, are not deformed as strongly
as cells migrate and squeeze. Hence, graphs for lower $\phi$ are compressed
horizontally. Just as the cellular shape diffusion relation in
\reffig{fig:shape-diffusion-law}[a-c], the curves in this nuclear shape diffusion
plot exhibit a transition between two slopes, with a crossover point that depends on
$\phi$.

The same stratified relation can be seen when plotting the
$\dmin$ from experiments against nuclear shape in 
\reffig{fig:shape-diffusion-law}[e-f]. These curves however are not
horizontally compressed, since we only consider a single system
with a single distribution of nuclear shapes and sizes. MCF-10A cells show larger mobility differences as they undergo a jamming
transition, while MDA-MB-436 cells remain unjammed (see
\ref{sec:experimental-data-aggregation}).

Remarkably, rescaling and shifting by nucleus-size-dependent parameters
(see sections \ref{methods:rescaling} and \ref{methods:analysis}) collapses all nuclear-shape graphs
onto single a curve
(\reffig{fig:shape-diffusion-law}[g-i]). Both simulations and experiments
therefore show clear evidence for our proposed mechanism of nuclear jamming,
since cell dynamics correlates directly with nuclear area fraction and shape.
Strikingly, despite representing two vastly different phenotypes, both
cell lines thus obey the same physical mechanisms that link nuclear morphologies
with collective dynamics. 

The collapsed curves again exhibit a crossover between two distinct slopes
(\reffig{fig:shape-diffusion-law}[g-i]), similar to the cell-shape-diffusion
relations (\reffig{fig:shape-diffusion-law}[a-c]). This resemblance shows that
nuclear shape affects cell mobility in the same way as cell shape, once we
properly account for the effects of $\phi$. Nuclear area fraction determines how
strongly nuclear and cellular shape are coupled: Larger nuclear area fractions
require that the nuclear shape more closely follows cell shape. Consequently,
the cell shape is more constrained by the nucleus in these cases.

The collapse establishes a novel relation between nuclear morphologies
and collective dynamics, i.e., a morpho-dynamic link. Inverting the rescaling procedure, we can predict the average
$\dmin$ from experiments, yielding a \gls{mape} (see \refeq{eq:mape} in section \ref{sec:shape-diffusion-exp}) of less than $5\,\%$ across both cell lines
(\reffig{fig:shape-diffusion-law}[g,i]). The
rescaled and shifted nuclear shape index on the x-axis of our shape-diffusion
relations (\reffig{fig:shape-diffusion-law}[g-i]) therefore acts as an order parameter
for nuclear jamming. This order parameter directly considers the control
parameters of nuclear jamming: stiffness (through nuclear shape) and density
(through the nuclear area fraction). The combination of nuclear shape and
area fraction strongly influences the cellular $\dmin$ in our experiments
(\reffig{fig:shape-diffusion-law}[l,m]). While cell shape can be influenced by
factors other than nuclear morphology, the nuclear shape diffusion relations in
\reffig{fig:shape-diffusion-law}[g-i] explicitly isolate the effects of nuclear
jamming on cell dynamics.

\section{Conclusion}
Our results establish nuclear area fraction and deformability as a key regulator
of collective dynamics in confluent cell monolayers. 
We reveal a universal relation between nuclear area fraction, stiffness, shape and diffusivity by introducing
deformable nuclei in a cellular Potts model of confluent 2D layers.
Our simulations qualitatively agree with in-vitro experiments
with MCF-10A and MDA-MB-436 cells, where smaller nuclear area
fractions and more elongated nuclei enhance mobility. After fitting cell
line specific parameters, our nuclear shape diffusion relation achieves a
predictive \gls{mape} below $5\,\%$ for both cell lines.

The model introduced in this work illuminates the mechanism by which
deformable nuclei affect tissue fluidity: Soft, small nuclei allow for
many T1 transitions, whereas rigid, large nuclei hinder tissue
reorganisation~\cite{jain_robust_2023, bi_density-independent_2015}. Consistent
with single-cell migration studies~\cite{wolf_physical_2013,
davidson2014nuclear, scianna_marco_cellular_2013, scianna_cellular_2021}, the
nucleus acts as a geometric obstacle. In densely-packed layers, it restricts
cell shape deformations and limits the ability of cells to rearrange, thereby
reducing collective mobility.

Importantly, our work unifies the previously opposing notions of
density-driven versus shape-driven (un)jamming~\cite{oswald_jamming_2017,
angelini_glass-like_2011, bi_density-independent_2015}. Nuclear stiffness and,
crucially, area fraction limit the range of accessible cell shapes in confluent
cell layers, ultimately determining the level of jamming. The
emergence of a universal shape-diffusion relation confirms cell shape as an
indicator for tissue fluidity. However, cell shape is ultimately constrained by
nuclear properties. The nucleus thus emerges as a pivotal mechanical
element mediating density effects and cell shape.

Our nuclear shape-diffusion relation links morphology and cell dynamics, offering a mechanistic bridge from nuclear deformation to collective
transport. Based entirely on statically observable nuclear properties, this
relation is particularly suitable for clinical settings, as it
could be analysed with established techniques and applied directly to
histological images. Nuclear grading is already part of the
histopathological evaluation of
tumours~\cite{bloomHistologicalGradingPrognosis1957}. Recently, changes in
nuclear morphology~\cite{gottheil_state_2023} and softening of cancer cells and
nuclei during \gls{emt}~\cite{fuhs_rigid_2022} have been associated with
unjamming and increased metastatic potential.
Our work strongly suggests a physical
mechanism underlies these empirical findings, with changes in nuclear
area fraction and stiffness contributing to an unjamming transition. After careful evaluation of the clinical context, the proposed
order parameter based on nuclear morphologies could complement existing markers as a
physically motivated morpho-dynamic link, potentially improving diagnostic
accuracy for cellular diseases such as cancer.

\bibliographystyle{apsrev4-1} 
\bibliography{./references}

\onecolumngrid

\section{Methods}

\subsection{Computational model}\label{methods:model}
We employ an extended \gls{cpm}~\cite{graner_simulation_1992} to simulate
confluent layers of cells with deformable nuclei. Inspired by the
compartmentalised description of single cells and their nuclei proposed by
\citet{scianna_cellular_2021}, we develop a model of confluent layers with
active cells. We thus extend the original \gls{cpm} by adding two new terms to
the Hamiltonian: (i) an active force and (ii) a restoring spring force,
anchoring the nucleus to the cytoplasm. Together with the original energy terms
introduced by \citet{graner_simulation_1992}, our full Hamiltonian for time step
$t$ reads
\begin{equation}
H^{(t)} = H_\mathrm{contact} + H_\mathrm{area} + H^{(t)}_\mathrm{active} + H_\mathrm{spring},
\end{equation}
with
\begin{alignat}{2}
    H_\mathrm{contact}           &= &\sum_{\langle i,j \rangle }    & J_{ij} (1 - \delta_{\sigma(i), \sigma(j)}) \\
    H_\mathrm{area}              &= &\sum_{\sigma \in \Omega}                  & \lambda_\sigma (A_\sigma - A_{\mathrm{t},\sigma})^2 \\
    H^{(t)}_\mathrm{active}      &= -&\sum_{\sigma \in \mathcal{C}} & \pmb{r}_\sigma \cdot \pmb{F}^{(t)}_\sigma \\
    H_\mathrm{spring}            &= -&\sum_{\sigma \in \mathcal{N}} & \frac{k}{2} (\pmb{r}_\sigma - \pmb{r}_{\Gamma(\sigma)})^2.
\end{alignat}
Here, $i$ and $j$ enumerate pixel identities, $\sigma$ is the object ID and
$\tau$ the object type function (evaluating to either cell or nucleus),
and $\Gamma$ is a function associating each cytoplasm with a
corresponding nucleus through a cell ID. The latter is the same for a cytoplasm
and the particular nucleus that belongs to it, thus labelling cells as a whole.
It is thus distinct from the object ID, which is different for each object (i.e.
each nucleus and each cytoplasm) in the simulation. $\mathcal{C}$ and
$\mathcal{N}$ denote the sets containing the object IDs of all cytoplasms and
nuclei, respectively, and $\Omega = \mathcal{C} \cup \mathcal{N}$ is the set of
all object IDs in the system. $A_\sigma$ and $A_{\mathrm{t},\sigma}$ are the
instantaneous and target area of object $\sigma$; $\pmb{r}_\sigma$ denotes the
\gls{com} of object $\sigma$ and $\pmb{r}_{\Gamma({\sigma})}$ the \gls{com} of the
whole cell $\Gamma(\sigma)$ (i.e., cytoplasm and nucleus), and $k$ is the spring
constant of the harmonic force anchoring the nucleus to the cell's \gls{com}.
Finally, $\pmb{F}_\sigma^{(t)}$ denotes an active force on the cytoplasm and is
further defined in \refeq{eq:active_force}.

The contact energy term describes the interplay between object-object adhesion and
membrane tension~\cite{magno_biophysical_2015} with
\begin{equation}
    J_{ij} = J_\mathrm{ext}(\tau(i), \tau(j))(1 - \delta_{\Gamma(i), \Gamma(j)}) + J_\mathrm{int}(\tau(i), \tau(j)) \delta_{\Gamma(i), \Gamma(j)}.
\end{equation}
We use the intracellular contact energy between cytoplasm and nucleus
$J_\mathrm{int}$ as a control parameter for nuclear stiffness, with larger
values of $J_\mathrm{int}$ leading to stiffer, more spherical nuclei.
In addition to that, we impose a large energy penalty (see table
\ref{tbl:j_ext}) as soon as a nucleus is in contact with the cytoplasm of a
foreign cell through $J_\mathrm{ext}$, keeping the nucleus within its
corresponding cell at all times.

$H_\mathrm{area}$ is the common, quadratic energy
contribution for deviations of the object area $A_\sigma$ from a target value, set to $A_{\mathrm{t}, \sigma} =
900\,\mathrm{px}^2$ for cells. Nuclei are subject to an analogous constraint with a
target area determined by the nuclear area fraction $\phi$. Both constraints are
multiplied with a strength modulator of $\lambda_\mathrm{area} = 0.1\,k_\mathrm{B}T /
\mathrm{px}^2$.

Cells have the ability to migrate by exerting forces on
their environment~\cite{lecuit_force_2011, anon_cell_2012}. $H_\mathrm{active}$
mimics this behaviour by applying an active Brownian
force~\cite{henkes_dense_2020} of magnitude $F_\mathrm{active} = 0.5\,k_\mathrm{B}T /
\mathrm{px}$ to each cytoplasm's centre of mass with
\begin{equation} \label{eq:active_force}
\pmb{F}^{(t)}_\sigma = F_\mathrm{active} (\cos \theta^{(t)}_\sigma, \sin \theta^{(t)}_\sigma)^T, \quad \sigma \in \mathcal{C}.
\end{equation}
The polarity $\theta$ of the active force diffuses with $D_\theta =
2\cdot10^{-3}\, \mathrm{MCS}^{-1}$ according to
\begin{equation}
\theta_\sigma^{(t+1)} = \theta_\sigma^{(t)} + \sqrt{2 D_\theta} \xi, \quad \sigma \in \mathcal{C},
\end{equation}
where $\xi$ is a random Gaussian number with zero mean and unit variance.

Force transmission between the cytoskeleton and the
nucleus~\cite{mcgregor_squish_2016} is modelled by including a harmonic spring
with zero rest length and spring constant $k = 0.2\,k_\mathrm{B}T / \mathrm{px}^2$, linking
the nucleus to the centre of mass of its corresponding cell by means of
$H_\mathrm{spring}$.

For each tested set of control parameters $\phi$ and $J_\mathrm{int}$ (i.e. each data point in
\reffig{fig:msd_phase_diagram}[c-e]) we equilibrate 10 independent realisations with 100
cells each for $5\cdot10^4\,\mathrm{MCS}$ and then run for
$5\cdot10^6\,\mathrm{MCS}$ on a $300\,\mathrm{px} \times 300\,\mathrm{px}$
domain. Simulations are carried out using the \texttt{CompuCell3D}
simulation framework~\cite{swat_multi-scale_2012} with a neighbour order of 4 for the contact energies.

\begin{table}[bth]
\centering
\caption{\textbf{Intercellular contact energies $J_\mathrm{ext}$.} All values are given in units of $k_\mathrm{B}T$.}
\label{tbl:j_ext}

\begin{tabular}{|*{3}{c|}}
    \cline{1-2}
    Cytoplasm        & 2.0 \\ \cline{1-3}
    Nucleus          & 10.0 & 100.0 \\ \hline
    Cell type $\tau$ & Cytoplasm & Nucleus \\ \hline
\end{tabular}
\end{table}

\subsection{Experimental Data}
\subsubsection{Cell Culture \& Microscopy} \label{methods:experimental-setup}

MCF-10A cells and MDA-MB-436 cells were cultured according to standard protocol
\cite{juergenDiss}. Cells were seeded into 24-well plates with a flat microscopy
bottom at a cell density  marginally lower than the intended initial density of
the experiment.  To enable the cells to completely adhere, the well plates were
incubated for the entire night.

The MCF-10A cell monolayers are then imaged with a ZEISS Axio Observer spinning
disc microscope using a nuclear vital stain of (0.2 $\mu$M SiR-DNA,
Spirochrome). The MDA-MB-436 cells were observed with a ZEISS LSM 980 microscope using SPY650-DNA Spirochrome staining. A ZEISS EC
Plan-Neofluar M27 10X/0.3 objective was used for both cell lines. Images were recorded every 10
minutes over a duration of 3 days. To reduce photo-toxicity the deep-red laser
($\lambda_\text{ex} = 638\,\mathrm{nm}$) was used. The experiment took place in
an incubation chamber equipped with a large reservoir of medium.

\subsubsection{Cell Tracking}
The nuclei in the microscopy images of the cell monolayer images are segmented
with \textit{Cellpose-SAM} \cite{pachitariu_cellpose-sam_2025}. Subsequently,
the cell outlines are approximated via a watershed-based algorithm with the
nuclei segmentations acting as markers. 
This yields two binary masks, one for the nuclei and one for the cells. Cell
tracking is performed by an algorithm that performs linking of objects in
adjacent frames via maximising spatial overlap.
Starting from two labelled, temporally adjacent cell masks, it is recorded with
which objects in the second mask each object overlaps and how big that overlap
is. These records are organised in a tabular data structure which is
subsequently sorted by overlap size. Finally, the rows of this data structure
are iterated starting from the largest overlaps. An object in the first frame is
linked to the corresponding object in the second frame if no previous linking
for the first object already exists in the linking map to ensure that larger
overlaps are always prioritized. This is based on the observation that cells in
the analysed monolayers do not change their size nor their shape substantially
in between adjacent time points.

Based on the cell tracks provided by the overlap tracking algorithm, 
the cells' mobilities are quantified by the $D^2_\text{min}$ motility measure
which measures a cell's non-affine movement with regard to its neighbours $\mathcal{N}_i$\cite{PhysRevLett.100.208302}:

\begin{equation}
\label{eq:d2min}
    D^2_{\text{min}, i} = \underset{\pmb{E}}{\text{min}} \left[\frac{1}{N} \sum_{j \in \mathcal{N}_i} \left(\pmb{r}_{ij}(t+T) - \pmb{E}_i \pmb{r}_{ij}(t) \right)^2 \right]
\end{equation} 

In equation \ref{eq:d2min} $\bm{r_{ij}}(t)$ is the displacement vector between
cell $i$ and its directly adjacent neighbouring cell $j$. Calculating the
$D^2_\text{min}$ amounts to finding the strain tensor $\bm{E}$ minimizing the
mean squared difference between displacement vectors after lag time $T$ and the
linearly transformed by $\bm{E}$ displacement vectors at present. The
minimization problem is solved with the \texttt{minimize} function of the
\texttt{SciPy}-libray \cite{2020SciPy-NMeth}. The steps of the cell tracking
pipeline described here are implemented in the python programming language and
orchestrated with the  \texttt{nextflow} \cite{DiTommaso2017} workflow manager.
The source code for the pipeline steps is open-source and available in
\href{https://github.com/lettlini/cellular-dynamics-nf-modules}{this GitHub
repository}.

\subsection{Analysis} \label{methods:analysis}

\subsubsection{T1 transition detection algorithm} \label{methods:t1}

We detect T1 transitions with a simple, custom algorithm. First, snapshots of
our simulations are converted into undirected graphs, where edges connect
neighbouring cells. Each edge is represented by a sorted pair of cell IDs of
neighbouring cells. T1 transitions are detected by comparing the set of edges of
two consecutive snapshots at times $t_0$ and $t_1$ (here $1000\,\mathrm{MCS}$
apart). The difference of the two sets of edges reveals the set of gained and
the set of lost edges between the two points in time. The set of edges that
remained unaltered is computed as the union of the set of edges at $t_0$ and
$t_1$.

T1 transitions are characterised by the loss of an edge and the appearance of a
new (perpendicular) edge in a neighbourhood of four cells. T1 events can therefore be
found from the set of lost, gained and unchanged edges between two snapshots by
looking for a gained edge $(A, C)$ and a lost edge $(B, D)$ while ensuring that
$(A, B)$, $(B, C)$, $(C, D)$, and $(A, D)$ remain neighbours between $t_0$ and
$t_1$. The fact that edges are represented as sorted tuples ensures that relabelling $A$, $B$, $C$, and $D$
would always yield the same result, so it is sufficient to check a single
ordering of the labels. The full implementation of this algorithm can be found
in the \texttt{t1\_detection} module of this repository:
\url{https://github.com/lhillma/Cell2Image}.

\subsection{Simulation data processing}
For each simulation run, we store nuclear and cellular \glspl{com} every 100
\glspl{mcs} and full lattice snapshots every 1000 \glspl{mcs}. \glspl{msd} are
computed from the respective \glspl{com} after accounting for periodic boundary
conditions. System drifts due to the periodic space is removed by subtracting
the collective \gls{com} in every time step.

The ensemble-averaged cellular and nuclear shape index is defined as
\begin{equation} 
\langle S \rangle  = \left\langle \frac{P}{\sqrt{A}} \right\rangle,
\end{equation}
where $A$ is the area and $P$ is the perimeter of either the cell or the nucleus. 
The cell and nuclear shapes are calculated across all cells and saved snapshots for each set of parameters.
While obtaining the cellular and nuclear area can be achieved by
straight-forward pixel counting, estimating the perimeter of a continuous shape
on a discrete lattice is more involved as simple edge counting is insufficient
for accurate shape measurements. Here, we employ a strategy for estimating the
perimeter explained by~\citet{magno_biophysical_2015}, which involves
convolution with a circular mask to consider neighbouring lattice sites up to a
given neighbour order. We follow the recommendation
in~\cite{magno_biophysical_2015} by choosing a neighbour order of 8 for our
perimeter estimate. Details on the precise implementation of this method can be
found in ref.~\cite{magno_biophysical_2015}.

The cells' \glspl{msd} are calculated from the trajectories of the \glspl{com}.
Minimum image convention is applied in a first step to obtain absolute
displacements from positions in a periodic space. For efficient computation of
the \glspl{msd}, we then compute the displacement autocorrelation function
leveraging the \gls{fft} as described in
ref.~\cite{allenComputerSimulationLiquids2017}. Finally, system averaged
diffusion constants are determined by fitting a linear regression to the
\gls{msd} at lag times $\tau \geq 10^5\,\mathrm{MCS}$.

Simulations were ran for all 121 combinations of $J_\mathrm{int}$ and
$\phi$ in \reffig{fig:msd_phase_diagram}[c]. Cellular and nuclear shapes are
averaged across all recorded snapshots and all cells per simulation run. The
diffusion constants for each simulation run are subsequently plotted against
the thus determined average shape values to examine the shape diffusion relation
as outlined in the following section.

\subsubsection{Shape-diffusion relation} \label{methods:rescaling}

\begin{figure}
    \centering
    \includegraphics[width=\linewidth]{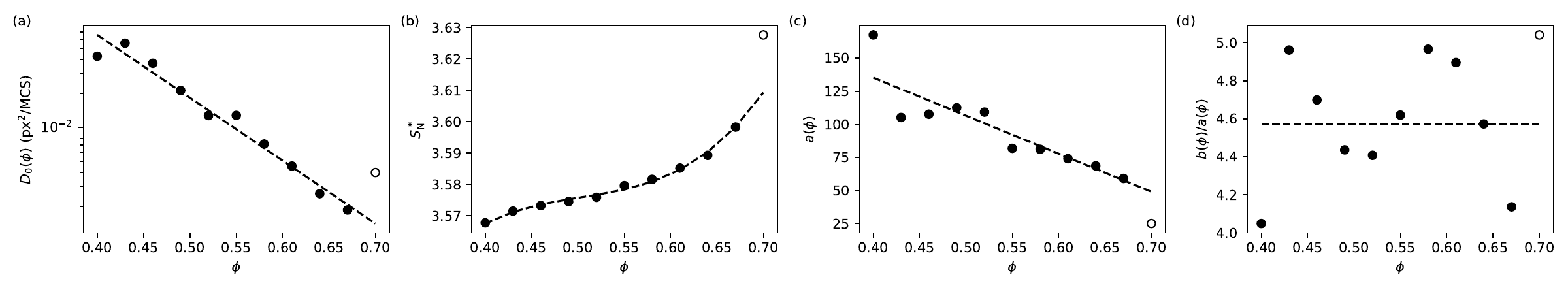}
    \caption{\textbf{Fitting procedure for the area-fraction-dependent
    parameters of the collapsed nuclear shape-diffusion relation.} \textbf{a},
    The collective diffusion constant decreases exponentially with increasing
    packing fraction. \textbf{b}, The critical nuclear shape index
    $S^*_\mathrm{N}(\phi)$ is well approximated by a third order polynomial.
    \textbf{c}, The initial slope of the shape-diffusion relation decreases
    approximately linearly with $\phi$. \textbf{d}, It is assumed that the ratio
    between slopes $a(\phi)$ and $b(\phi)$ remains approximately constant.}
    \label{fig:parameter_fits_phi}
\end{figure}

As discussed in the results, the plot of the average cell shape against the
average diffusion constant on semi-logarithmic axes exhibits two distinct
slopes. For our simulation data, we find that fitting the functional form
\begin{equation}
  D(S) = D_0 \left(e^{-a (S - S^*_\mathrm{C})} + e^{-b (S - S^*_\mathrm{C})}\right)^{-1}
  \label{eq:d_over_s}
\end{equation}
yields a stable and accurate fit of the two slopes $a$ and $b$ as well as the
transition shape index $S^*_\mathrm{C}$ and baseline diffusion $D_0$. Plotting $D$ against the nuclear shape
$S_\mathrm{N}$, there is a clear separation into distinct graphs for each
value of the area fraction $\phi$ (\reffig{fig:shape-diffusion-law}[d-f]). In
order to make a prediction of the diffusion coefficient of a tissue, it is
therefore required to know both $S_\mathrm{N}$ and $\phi$. An empirical law
for $D(S_\mathrm{N}, \phi)$ can be found by first fitting \refeq{eq:d_over_s}
for each value of $\phi$ separately and then considering the fit parameters to
be functions of $\phi$, effectively assuming
\begin{equation}
    D(S_\mathrm{N}, \phi) = D_0(\phi) \left(e^{-a(\phi) (S - S^*_\mathrm{N}(\phi))} + e^{-b(\phi) (S - S^*_\mathrm{N}(\phi))}\right)^{-1}. \label{eq:nucleus_diffusion_shape_}
\end{equation}

\reffig{fig:parameter_fits_phi}[a] suggests that the diffusion coefficient
decreases exponentially as the nucleus occupies more area in the cell. Since in
our simulations the nucleus is a stiff compartment within the cells, a general
decrease with increasing $\phi$ is consistent with previous observations in
comparable systems. Among others, simulations and experiments with Brownian hard
disks show that the diffusion coefficient decreases for larger packing fractions
of the disks~\cite{stopper_bulk_2018}. In contrast to~\cite{stopper_bulk_2018},
our data exhibit a linear decline in a logarithmic plot. Hence we chose to fit
\begin{equation}
    D_0(\phi) = \alpha \exp \left(\beta\phi\right)
\end{equation}
with fit parameters $\alpha$ and $\beta$.

Assuming that the underlying mechanism connecting shape fluctuations to
collective diffusion remains the same regardless of the nuclear packing
fraction, it is reasonable to expect that the ratio $\gamma$ between the slow rate
$a(\phi)$ and the fast rate $b(\phi)$ does not change:
\begin{equation}
    \frac{b(\phi)}{a(\phi)} = \gamma = \mathrm{const.}
\end{equation}
Although \reffig{fig:parameter_fits_phi}[d] shows deviations of
$b(\phi) / a(\phi)$ from a constant function, setting $\gamma = \mathrm{const.}$
proves to be sufficient to collapse the curves for different area fractions in
\reffig{fig:shape-diffusion-law}.

Plotting the slope $a(\phi)$ over the nuclear area fraction in
\reffig{fig:parameter_fits_phi}[c], we observe a linear trend with negative
slope. Hence, there are limitations at $a(\phi^*) = 0$, which effectively give
rise to an upper limit $\phi^*$ of the area fraction. The parameter $a(\phi)$ is
proportional to the rates in the two-step exponential. Considering this, one
notices that as $a$ becomes negative, the very nature of the diffusion shape
relation would change, because $a$ and $b$ would undergo a change of sign. Thus,
$a(\phi^*) = 0$ sets a formal upper limit for the area fraction. In our model,
the nucleus faces a high energy penalty once it is in contact with any cytoplasm
other than that of its own cell, which also gives rise to an upper limit of
$\phi$. Therefore, there has to be a gap of at least one pixel on either side of
the nucleus. The average cell diameter in our simulations is $30$ pixels, hence
$\phi^*$ can be estimated as
\begin{equation}
    \phi^* = \frac{28^2}{30^2} \approx 0.8711.
\end{equation}
Based on these arguments, we impose
\begin{equation}
    a(\phi) = \rho(\phi - \phi^*) = \rho\left(\phi - \frac{28^2}{30^2}\right)
\end{equation}
introducing the fit parameter $\rho$, which yields good agreement in \reffig{fig:parameter_fits_phi}[c].

Finally, the critical shape index $S^*(\phi)$ determines the transition point
between the two slopes in the double exponential law. We find
$S^*_\mathrm{N}(\phi)$ to be the most critical parameter in collapsing the
data in \reffig{fig:shape-diffusion-law}. From the plot in
\reffig{fig:parameter_fits_phi}[b], $S^*_\mathrm{N}(\phi)$ appears to have an
inflection point close to $\phi = 0.5$. The lowest order polynomial that can fit
this behaviour is a cubic function. In addition, note that at $\phi = \phi^*$
the nucleus fills the entire cell, and thus its shape index is close to the cell
shape index. We therefore demand $S^*_\mathrm{N}(\phi^*) = S^*_\mathrm{cyt}$,
which is obtained from the fit in \reffig{fig:shape-diffusion-law}[a]. Thus, we choose
\begin{equation}
    S^*_\mathrm{N}(\phi) = \kappa \left[(\phi - \mu)^3 - (\phi^* - \mu)^3\right] + \nu (\phi - \phi^*) + S^*_\mathrm{cyt}
\end{equation}
with parmeters $\kappa$, $\mu$, and $\nu$.

\begin{table}[h]
    \centering
    \begin{minipage}{.45\linewidth}
    \caption{Fit parameters obtained from plotting the collective diffusion coefficient over the cell shape.}
    \begin{tabular}{lcccc}
        \toprule
        Parameter &   $a$   &   $b$   & $S^*_\mathrm{C}$ & $D_0$ \\
        \midrule
        Value     &    34.9 &    3.04 &                3.75 & -5.75 \\
        \bottomrule
    \end{tabular}
    \end{minipage}
    \hfill{}
    \begin{minipage}{.45\linewidth}
    \caption{Fit parameters obtained from plotting the collective diffusion coefficient over the nuclear shape.}
    \begin{tabular}{lccccccc}
        \toprule
        Parameter & $\alpha$ & $\beta$ & $\gamma$ & $\rho$ & $\kappa$ & $\mu$ &      $\nu$          \\
        \midrule
        Value     &   11.1  &   -12.8 &     4.57 &   -287 &     3.21 & 0.503 & $4.63 \cdot 10^{-2}$ \\
        \bottomrule
    \end{tabular}
    \end{minipage}
\end{table}

\subsubsection{Experimental data aggregation \& shape diffusion relation}
\label{sec:experimental-data-aggregation}

In contrast to our simulations, there is natural variability in biological
tissues. Furthermore, while our simulations operate at a fixed nuclear packing
density, packing density increases over time in our experiments with MCF-10A cells. Cells
and nuclei also exhibit natural size polydispersities, so, packing
density also varies spatially. Therefore, cells and nuclei in a single frame
have different physical properties and different regions in the same
image exhibit varying degrees of jamming and unjamming. Due to these spatial
variations within a frame, the \emph{frame-averaged} $\dmin$ is not a suitable
measure of average cell mobility. Instead, we group nuclei and cells by dividing
our data into equally-sized bins for $\phi$ and cellular (nuclear) shape. We
choose 20 bins for nuclear packing density and 40 bins for the shape index,
respectively. For each of these bins, we then compute the average $\dmin$ of the
cells belonging to that bin and only consider bins that contain at least 100
observations.

This way of grouping cell observations by the observed packing fraction and
shape index is consistent with the aggregation by the imposed target
packing density and surface tension $J_\mathrm{int}$ of the simulation data.
However, binning based on measured properties changes the scales of the axes
of the shape diffusion law plots due to fluctuations. More precisely, compared
to the data aggregation of the simulation data, there is a larger range of
shape indices, because the instantaneous shape index is subject to larger
fluctuations than time-averaged shape. The same holds for the nuclear packing
density. Conversely, the range of average $\dmin$ decreases compared
to our analysis of simulated data, because instantaneous shape and density
fluctuations cause cells with a specific $\dmin$ to be distributed over
multiple bins. Thus, the distribution of $\dmin$ within the bins become
broader and the averages of adjacent bins move closer together.

\subsubsection{Experimental shape diffusion relation} \label{sec:shape-diffusion-exp}
The averaged cell $\dmin$ per bin are then plotted against each bin's
average shape index, as described in section~\ref{methods:rescaling}.
However, we find that eq.~(\ref{eq:d_over_s}) does not yield a stable fit in
our experiments. Therefore, consistent with the claim that we observe two different
slopes on a logarithmic plot, we instead fit the piece-wise function

\begin{equation}
    \log{D(S, \phi)} = D_0(\phi) + \begin{cases}
        a(\phi) (S - S^*(\phi)) & S < S^*(\phi) \\
        b(\phi) (S - S^*(\phi)) & \mathrm{otherwise.} \\
    \end{cases}
    \label{eq:exp-shape-diffusion-relation}
\end{equation}

Just as above, we can again assert that $\frac{a}{b} = \gamma = \mathrm{const.}$
and assume a linear relation for $a(\phi)$. In contrast to our simulations, here
there is no simple argument for an upper limit for $\phi$. We therefore impose a
general linear function

\begin{equation}
    a(\phi) = \rho(\phi - \sigma).
\end{equation}

Finally, we observe that from visual inspection the function $S^*(\phi)$ follows
a linear, rather than a cubic relation, which is consistent with the above,
since it is a special case of the third order polynomial used there. 

Although our fitting procedure for the experimental data thus differs
from our analysis of cell dynamics in simulations in these three aspects, all
modifications are still consistent with our previous rescaling method (see
section~\ref{methods:rescaling}). Considering the challenges of comparing
inherently polydisperse and heterogeneous cell shapes and trajectories from
in-vitro experiments with monodisperse simulations, it would be expected that
some of the underlying assumptions have to change. Despite these inherent discrepancies
between the simulated and experimental system, we surprisingly still observe the
same qualitative relation between cellular and nuclear morphologies and mobility.
Once the fit parameter in \refeq{eq:exp-shape-diffusion-relation} are
determined, our model can be used to predict the expected $\dmin$ at a given
nuclear area fraction and shape index. We measure the accuracy of the predicted
$\dmin$ from our model using the \acrfull{mape}, defined as
\begin{equation}
    \mathrm{MAPE} = \frac{1}{N} \sum_{i=1}^N \left | \frac{y_i - \hat{y}_i}{y_i}\right |,
    \label{eq:mape}
\end{equation}
where $y_i$ and $\hat{y}_i$ are measured $\dmin$ and the $\dmin$ predicted by
our model for the $i$-th bin, respectively, and $N$ is the total number of bins.

\section{Acknowledgements}

We thank Corentin Laudicina and Ilian Pihlajamaa for useful discussions and their critical reading of the manuscript. 
L.H., S.H., M.V., and L.M.C.J.\ gratefully acknowledge 
the Eindhoven Artificial Intelligence Systems Institute (EAISI) for funding through the EMDAIR program. 
J.A.K.\ acknowledges funding under Research Unit FOR5628 granted by the Deutsche
Forschungsgemeinschaft (DFG). 
Q.J.S.B.\ and L.M.C.J.\ thank the Dutch Research Council (NWO) for financial support through the ENW-XL project ``Active Matter Physics of Collective Metastasis" (OCENW.GROOT.2019.022).

\end{document}